\documentstyle[12pt]{article}
\newcommand{\tr}[1]{\,{\rm tr}\,#1\,}

\begin{document}
\def\newmathop#1{\mathop{#1}\limits}
\def\ninfty#1{\displaystyle\newmathop{#1}_{N\to\infty}}
\title{
\begin{flushright}
{\small SMI-5-93 \\ May, 1993 }
\end{flushright}
\vspace{2cm}
Large N QCD at High Energies\\ as\\Two-Dimensional Field Theory}
\author{I.Ya. Aref'eva \thanks{E-MAIL: Arefeva@qft.mian.su}
\\ Steklov Mathematical Institute,\\ Russian Academy of Sciences,\\
Vavilov st.42, GSP-1,117966, \\ Moscow, Russia }
\date{~}
\maketitle
\begin {abstract}

Different aspects of the Verlinde and Verlinde
relation between high-energy effective scattering in QCD
and a two-dimensional sigma-model are discussed.
Starting from a lattice version
of the truncated 4-dimensional Yang-Mills action we derive an effective
theory with non-trivial longitudinal dynamics which has a form of the
lattice two-dimensional chiral field model
with non-trivial boundary conditions. To get quantum corrections coming
from non-trivial longitudinal dynamics to transversal high-energy
effective action one has to solve the two-dimensional chiral field model
with non-trivial boundary conditions.  We do this within an approximation
scheme which takes into account one-dimensional excitations.
Contributions of the one-dimensional excitations  to quantum corrections
for the high-energy effective action are calculated  in the large N limit
using the character expansion method.

\end {abstract}

\newpage

\section{Introduction}

One of the striking results of intensive studies of high-energy
scattering in
QCD \cite {LipR,ChWu} is that the scattering amplitudes are related with
two-dimensional field theory \cite {Lip}.
Recently H.Verlinde and E.Verlinde have formulated a simple model
in which this two-dimensional nature of the interactions is
manifest \cite {VV}. They  have also  shown that their formulation is in
agreement with known results from standard perturbation theory  \cite
{LipR,ChWu}.

There are some questions concerning the Verlinde and Verlinde
approach which we would like to address in this paper. First question is
related with ultra-violet divergences and renormalizations. The Verlinde
and Verlinde effective action which describes the transversal dynamics has
a form of the two-dimensional $\sigma$-model. In spite of a non-convention
form  of this $\sigma$-model it is reasonable to expect that it is
asymptotic free. But the $\beta $-functions corresponding to the
4-dimensional Yang-Mills theory  and the two-dimensional  $\sigma$-model
numerically have the different forms.  Namely, the 4-dimensional
Yang-Mills  $\beta $-function  contains the factor $\pi$ and
$\beta $-function of the two-dimensional $\sigma$-model
does not. So one can expect that an additional renormalization
comes from the longitudinal dynamics.
This circumstance give us a raison to study
the longitudinal dynamics more carefully. A treatment of
longitudinal dynamics within the framework of the
initial truncated Yang-Mills action is a main goal of this paper.

 To analyze quantum corrections to  the longitudinal dynamics  and the
corresponding  renormalization we have to
introduce some regularization. As a regularization we can introduce a lattice
in x-space. Recall that  one can deal with 4-dimensional lattice only in
the Euclidean space-time. However, the high-energy effective action has
been obtained  \cite {VV} in the Minkowski space-time.  The Minkowski
signature is essential, since it permits to use the property of the
two-dimensional wave equation. We would like to know does a similar action
arise in the Euclidean formulation and what are its quantum corrections.
For this purpose we study the lattice version of the truncated action.
It turns out that the lattice version of the truncated action looks like
the usual lattice version of the two-dimensional chiral model
with non-trivial boundary conditions. An interaction between fields living on
the transversal planes appears due to the boundary effects in the
two-dimensional chiral model. The lattice two-dimensional chiral model
with non-trivial boundary conditions  cannot be solved exactly.
One can consider the one-dimensional chiral model
and analyze the form of the
corresponding transversal  effective action. As in the two-dimensional
case we can speculate that the one-dimensional
answer describes the contribution coming from long
one-dimensional  excitations. In other words one can estimate
long tubes contribution to the quantum version ofVV effective action.

The paper is organized as follows.
In section 2 we present a brief review of the Verlinde and Verlinde
results and write down a lattice version
of the truncated action. In section 3 we study the longitudinal dynamics for
a model case then this dynamics is one-dimensional and calculate a leading
term of the high-energy effective action for large N. We discuss the
two-dimensional  longitudinal dynamics in a concluding section.

\newpage

\section{Lattice Effective Action for QCD at High Energies.}
\subsection{Truncated Action}
Since this paper is inspired by the recent H.Verlinde and E.Verlinde  paper
\cite {VV}, let us start from a brief presentation of their results. They are
interested in the behaviour of scattering amplitudes  in  the
kinematical regime where $s$ is much larger than $t$, while
$t$ is also larger than the QCD scale $\Lambda_{qcd}$.
Introducing two light-cone
coordinates $
x^\alpha = (x^+,x^-) $ with $x^\pm = x \pm t$  and two transverse coordinates
$ z^i = (y, z)$, authors assume that two fast moving
particles  have very
large momenta in the $x^\pm$ direction, while they
remain at a relatively large distance in the $z$-direction.
They also use the hypothesis, according to which the typical longitudinal
momentum of the dynamical modes in this process grows proportionaly
to the centre of mass energy, whereas the typical size of the
transversal momenta is determined by the momentum transfer.
A similar assumption was used also in other approaches to high
energy QCD  \cite {LipR}. The distinguished feature of the H.Verlinde and
E.Verlinde approach is that the high-energy limit is taken directly at the
level of the action.

Let us do this for the pure Yang-Mills model, described by
\begin {equation} 
							  \label {1.1}
S = \frac{1}{4g^{2}} \int d^4x \tr (F_{\mu\nu}F^{\mu\nu}).
\end   {equation} 
Here $F_{\mu\nu}$ is the non-abelian field strength,
$ F_{\mu\nu} =
\partial _\mu A_\nu - \partial_\nu A_\mu + [A_\mu,A_\nu]$ ,
$A_\mu = A_\mu^a \tau^a $
where $g$ is the coupling constant and
$\tau^a$ are the generators of the Lie algebra of the
gauge group $G = SU(N)$.

The main idea  \cite {VV} is that by  performing a rescaling of the
longitudinal coordinates
$
x^\alpha \to  \lambda x^\alpha $
inside the Yang-Mills action (\ref {1.1}), one can see which part will
become strongly or weakly coupled when one looks at the theory at high
longitudinal energies.
The components of the gauge potential transform under this rescaling as
$A_i \to A_i$, while $A_\alpha \to \lambda^{-1} A_\alpha$ and the rescaled
Yang-Mills action can be written as
\begin {equation} 
							  \label {1.1'}
S_{YM}^\prime = \frac{1}{4\lambda^{2} g^{2}} \int \tr ( F^{\alpha\beta}
F_{\alpha\beta} )
+\frac{1}{2g^{2}} \int \tr ( F^{\alpha i}F_{\alpha i} )
+\frac{\lambda^{2}}{4g^{2}}\int \tr(F_{ij}F^{ij}).
\end   {equation} 
Introducing an auxiliary field	$E_{{\alpha\beta}}=-E_{\beta\alpha}$ one can
rewrite (\ref {1.1'}) in the form
\begin {equation} 
							  \label {1.2}
S_{YM}^\prime  = \frac{1}{2g^{2}} \int \tr ( E^{{\alpha\beta}}F_{\alpha\beta}
+F_{\alpha i}F^{\alpha i}) + \frac{\lambda^2}{4g^{2}}\lambda^2
\int \tr(E_{{\alpha\beta}}E^{{\alpha\beta}}+F_{ij}F^{ij}).
\end   {equation} 
The description of a scattering process with some $s$ and $t$ using
the standard action is completely equivalent to that using the
rescaled action (\ref{1.2}) with  rescaled $s$ to
$
s'= \lambda^2 s$, i.e.
\begin {equation} 
							  \label {1.3}
{\cal A}(S_{YM},s) ={\cal A}(S_{YM}^\prime ,\lambda^2 s),
\end   {equation} 
where ${\cal A}$ is a scattering amplitude describing the process.
One can use this correspondence
and reformulate the high-energy
limit $s\to \infty$ in QCD as the   $\lambda \to 0$
limit of the rescaled theory (\ref{1.2}) with $s'$ fixed, i.e.
$\lambda \sim {1\over \sqrt{s}} \to 0$.

Therefore, high-energy scattering in gauge theories can be studied by means
of the truncated Yang-Mills action
\begin {equation} 
							  \label {1.4}
S[A,E] ={1\over 2 g^{2}}\int \tr(E^{\alpha\beta }F_{\alpha\beta }
+ F_{\alpha i}F^{\alpha i} ).
\end   {equation} 
The auxiliary field $E_{\alpha\beta }$	 can be integrated out
yielding an effective theory describing the zero-curvature
longitudinal gauge-field $A_\alpha$  of the form
\begin {equation} 
							  \label {1.5}
A_\alpha= \partial_\alpha V V^{+}.
\end   {equation} 
Here the group element $V$ may still depend on all four coordinates
\footnote{Jacobian of this replacement of the variables $E^{\alpha\beta}$ and
$A_\alpha$ by $U$ is simply equal to 1	\cite {VV} }.
Substituting (\ref{1.5}) into the action $S$, one gets
\begin {equation} 
							  \label {1.6}
S[V,A_i]
= {1\over 2 g^2}\int\, \tr (\partial_\alpha (V^{+}D_iV))^2,~~~
D_i  = \partial_ i  +  A_i .
\end   {equation} 
At the formal level the longitudinal   components $A_\alpha$ are completely
eliminated by a gauge transformation of $A_i$ to
$\widetilde{A}_i = V^{-1}D_i V$, i.e.
\begin {equation} 
							  \label {1.7}
S[V,A_i]
= {1\over 2 g^2}\int ~ \tr (\partial_\alpha \widetilde{A}_i)^2.
\end   {equation} 
However if we consider an interaction of gauge fields with charged particles,
then non-arbitrary gauge transformations at infinity are allowed.
\subsection{Lattice Truncated Action}

Another subtle question related to the limit $\lambda \to 0$ is the
problem of divergences. To avoid ultra-violet divergences one can put
the theory on the lattice.
Let us consider a lattice version of the rescaled action (\ref {1.1'})
\begin {equation} 
							  \label {1.8}
S'_{LYM}=\frac{1}{4\lambda ^{2}g^{2}}\sum _{x}\sum _{\alpha,\beta }
(\tr U(\Box _{\alpha,\beta })-1) +
\frac{1}{4g^{2}}\sum _{x}\sum _{\alpha, i }
(\tr U(\Box _{\alpha, i })-1)
\end   {equation} 
$$+\frac{\lambda ^{2}}{4g^{2}}\sum _{x}\sum _{i,j }
(\tr U(\Box _{i,j})-1).
$$
Here $x$ are the
points of the 4-dimensional lattice, $\alpha,\beta $ are unit
vectors in the longitudinal direction and $i,j$ are unit vectors in
the transversal direction; $\Box _{\mu , \nu }$  is a single plaquette
attached to the links $(x,x+\mu)$ and $(x, x+\nu)$, $\mu ,\nu =\alpha ,i$.
Performing
the $\lambda \to 0$ limit in the rescaled lattice theory (\ref {1.8})
we get a lattice version of the truncated Lagrangian (\ref {1.6})
\begin {equation} 
							  \label {1.9}
S'_{LYM}=\frac{1}{4g^{2}}\sum _{x}\sum _{\alpha, i }
\tr (U(\Box _{\alpha, i }) -1),
\end   {equation} 
with $U_{x,\alpha}$ being a subject of the relation
\begin {equation} 
							  \label {1.10}
U(\Box _{\alpha,\beta })-1 =0,
\end   {equation} 
i.e. $U _{x,\alpha }$ is a zero-curvature lattice gauge field
\begin {equation} 
							  \label {1.11}
U _{x,\alpha }=V _{x}V ^{+}_{x+\alpha }.
\end   {equation} 
Substituting (\ref {1.11}) in (\ref {1.9}) we get
\begin {equation} 
							  \label {1.12}
S_{tr}=\frac{1}{4g^{2}}\sum _{x}\sum _{\alpha, i }
\tr (V_{x+\alpha}^{+}U_{x+\alpha,i}V_{x+\alpha +i}
V^{+}_{x+i}U^{+}_{x, i }V_{x}-1).
\end   {equation} 

Eliminating $V _{x}$ by the gauge transformation
$U_{x, i } \to \tilde{U}_{x, i }=V^{+}_{x}U^{+}_{x, i }V_{x+i}$  we get
the lattice version of the Lagrangian (\ref {1.8}), i.e.
\begin {equation} 
							  \label {1.13}
S_{tr}=\frac{a^{D-4}}{4g^{2}}\sum _{x}\sum _{\alpha, i }
\tr (U_{x,i}U^{+}_{x+\alpha, i }-1).
\end   {equation} 
Let us denote the points of 4-dimensional lattice as $x=(y,z)$, where $y$ and
$z$
are the points of two two-dimensional lattices, say, y-lattice (longitudinal)
and z-lattice (transversal).
We see that the action (\ref {1.13}) has a factorisation form
\begin {equation} 
							  \label {1.13a}
S_{tr}=\frac{a^{D-4}}{4g^{2}}\sum _{i,z}
\sum _{y,\alpha }\tr (U_{y,z;i}U^{+}_{y+\alpha,z;i }-1).
\end   {equation} 
Here we restore for the
moment a factor $a^{D-4}$, which is omitted in the previous
formula since we assumed $D=4$. The indices $z$ and $i $ play a role of
isotopic indices.
Let $(L)^{2}$ is a number of points  of  the z-lattice. Then we can say that
we have $2(L)^{2}$-copies of
two-dimensional lattice  models. Each model  describes
the chiral field $ U_{y}$,  attached  to the sites of the two-dimensional
$y$-lattice
\begin {equation} 
							  \label {1.13b}
\tilde{S}_{tr}=\frac{1}{4g^{2}}\sum _{y,\alpha }
\tr (U_{y}U^{+}_{y+\alpha }-1).
\end   {equation} 
If one forgets about subtlety with the boundary conditions then for trivial
boundary conditions  one can say that the free energy per unit volume
 of the model (\ref {1.13}) can be represented as
\begin {equation} 
							  \label {1.x}
E(N,g(a),a )=\frac{1}{Vol _{4}}\ln [\int \exp (S_{tr})\prod _{y,z,i}
dU_{y,z,i}] =
\frac{2}{Vol _{2}}\ln [\int \exp (\tilde{S}_{tr})\prod _{y}
dU_{y}].
\end   {equation} 

Note that there is some similarity between $\lambda$
rescaling and the
high temperature limit of lattice gauge theory \cite{Dadda2},
in the first case one
makes a decomposition $(2,d-2)$ and in the second $(1,d-1)$.

 The classical continuous limit of the action (\ref {1.13b}) gives the usual
two-dimensional chiral model
\begin {equation} 
							  \label {1.ca}
S_{tr, cl}=\frac{1}{4g^{2}}\int d^{2}x
\tr (\partial _{\alpha}U(x)\partial _{\alpha}U^{+}(x)).
\end   {equation} 

Therefore,
removing two-dimensional lattice by  $a\to 0$  limit in (\ref {1.13})
one gets $(L)^{2(D-2)}$-copies of the two-dimensional continuous models.

To remove the $y$-lattice in the quantum version of the  two-dimensional
chiral model
one has to perform the renormalization. In the one-loop approximation
\cite {Ren} one has
\begin {equation} 
							  \label {1.14}
\frac{1}{g^{2}(a)} =\frac{1}{g^{2}(\mu)} +c_{ch}\log (a\mu )^{1}
\end   {equation} 
with
\begin {equation} 
							  \label {1.16}
c_{ch}= N.
\end   {equation} 

Note that to remove a lattice in the regularized version of the quantized
 Yang-Mills  theories  one has to perform the renormalization
\begin {equation} 
							  \label {1.15}
\frac{1}{g^{2}(a)} =\frac{1}{g^{2}(\mu)} +c_{YM}\log (a\mu )^{1}
\end   {equation} 
with
\begin {equation} 
							  \label {1.16a}
c_{YM}= \frac{1}{16\pi ^{2}}\frac{11N}{3}.
\end   {equation} 

Both theories are asymptotically free, but the renormalizations of coupling
constants do not coincide. This means that taking the limit $\lambda \to 0$
we drop diagrams which contribute to the renormalization of the coupling
constant.

 Note also that (\ref {1.13}) does not give a usual
4-dimensional  model in the limit $a\to 0$ , since we have
not an extra factor $(a)^{2}$ which would lead us to 4-dimensional continuous
theory.

Let us	compare the Lagrangian (\ref {1.13}) with the Lagrangian
(\ref {1.7}).  Assuming that $U_{x, i}$ has a form  $U_{x, i}=
\exp aA_{i}(x_{\pm},z)$
we reproduce  (\ref {1.7}),
since extra factor $a^{2}$ together with summation
over the z-lattice produces the integral over the  z-plane. At first sight
there are no raisons
to get propagator modes in the z-plane, since there are not the corresponding
kinetic terms. The Verlinde and Verlinde have observed that a non-trivial
dynamics  in the transversal z-plane  can be introduced due to non-trivial
boundary conditions in the $x_{\pm}$-plane. Namely,
they rewrote the action (\ref {1.6})   as
\begin {equation} 
							  \label {1.vva}
S[U,A_i^{cl}] = {1\over 2 e^2} \int \! d^2 z \, \tr [
\int^{{}^{\infty}}_{{}_{-\infty}} \!\!\!\!\!\! dx^+
\, \partial_+ (U^{-1}D_iU)
\, \times \,
\int^{{}^{\infty}}_{{}_{-\infty}} \!\!\!\!\!\! dx^-\, \partial_-
(U^{-1}D_i U) ]
\end   {equation} 
and then  expressed the right-hand side directly in terms of the
asymptotic values for $U$ and $A_i$ as follows
$$
\, \int^{{}^{\infty}}_{{}_{-\infty}} dx^-\, \partial_- (U^{-1}D_iU) =
g_{2}^{+}D^+_i g_{2} - g_{1}^{+}D^+_ig_{1},$$
$$
\int^{{}^{\infty}}_{{}_{-\infty}}
dx^+\, \partial_+ (U^{-1}D_iU)~=~h_{2}^{-1}D^-_i h_{2}-h^{-1}D^{-}_{i}h_{1},
$$
where the covariant derivatives $D^\pm_i$ on the
right-hand sides are with respect to $a^{\pm}_{i}$.
The effective action of the two-dimensional variables
$(g_{A},h_{B},A_i^\pm)$ they represented as
\begin {equation} 
							  \label {1.ea}
S[g_{A},h_{B},A_i^\pm] = {1\over 2g^2} \int d^2 z
M^{AB} \tr(g_{A}^{+}D^{+}_i g_{A}   h_{B}^{+}D^-_i h_{B} ),
\end   {equation} 
where $M^{AB}$ is the $2\times 2$ matrix,
$M^{AB} = \left(\begin{array}{cc} 1 & -1 \\
-1 & 1 \end{array}\right)$
and the indices {\small{$A$}} ({\small {$B$}}) $=1,2$ are summed over.
In  \cite {VV} was argued that the matrix $M^{AB}$ should be regularized
as
\begin {equation} 
							  \label {1.r}
M_{reg}^{AB} = \left(\begin{array}{cc} 1+\epsilon & -1+\epsilon \\
-1+\epsilon & 1+\epsilon \end{array}\right),
\end   {equation} 
where
$$\epsilon ^{-1}=1-\frac{2i}{\pi}\log s .$$
Note that to get the representation (\ref {1.ea}) one has to
deal with the Minkowski signature. Working with the usual lattice
approximation, i.e. dealing with the Euclidean signature one cannot get in
the classical approximation an analog of the representation (\ref {1.ea})
typical for solutions of the wave equation.

\section{One-dimensional Longitudinal Dynamics}

Now let us discuss effects related with non-trivial boundary conditions in
the action (\ref {1.13a}). It is rather instructive to consider a model example
of the one-dimensional y-lattice. So, we deal with the following lattice
integrals
\begin {equation} 
							  \label {m.1}
Z_{1+2}=\int \exp \{ \frac{aN}{4g^{2}}\sum _{i,z}
\sum _{y}\tr (\tilde{U}_{y,z;i}\tilde{U}^{+}_{y+1,z;i }
+\tilde{U}_{y+1,z;i}\tilde{U}^{+}_{y,z;i }
-2)\}
\prod _{i,z,y}dU_{y,z;i}dV_{z;i}
.\end	{equation} 
More precisely we mean the following boundary conditions
\begin {equation} 
							  \label {m.2}
Z_{1+2}=\int \prod _{(z,i)}(\int \exp \{ \frac{aN}{4g^{2}}
\sum _{y}\tr (\tilde{U}_{y,z;i}\tilde{U}^{+}_{y+1,z;i }
+\tilde{U}_{y+1,z;i}\tilde{U}^{+}_{y,z;i }
-2)\}.
\end   {equation} 
$$\times\prod _{\begin{array}{c}\tilde{U}_{0,z;i}
=V_{0,z}U_{0,z;i}V^{+}_{0,z+i}\\
\tilde{U}_{L,z;i}=V_{L,z}U_{L,z;i}V^{+}_{L,z+i}\end {array}
}d\tilde{U}_{y,z;i}dU_{0,z;i}dU_{L,z;i}dV_{0,z}dV_{L,z}
$$
In (\ref {m.2}) we have the product of the transition functions of the
one-dimensional lattice chiral model. This correlation function can be
calculated \cite {Sigma}  using the characters expansion method,
\begin {equation} 
							  \label {m.3}
{\cal K}(U_{0},U_{L}) =
\int \exp \{N\tilde{\beta}\sum _{n=0}^{L}\tr (U_{n}U^{+}_{n+1}
+U_{n+1}U^{+}_{n}-2) \}\prod _{n=1}^{L-1}dU_{n}
\end   {equation} 
$$
=\sum _{R}dim_{R}(Z_{R})^{(L-1)}\chi _{R}(U_{0}U^{+}_{L}),
$$
where $\chi _{R}$ and $dim_{R}=\chi _{R}(I)$ are characters and dimensions of
the R's irreducible representation respectively.
Here $L'=L-1$.
$Z_{R}$ are given by the following formula
\begin {equation} 
							  \label {m.4}
Z_{R}=\frac{1}{dim_{R}}\int dU \chi _{R}(U^{+})\exp \{N\tilde {\beta}
\tr (U+U^{+}-2)\},
\end   {equation} 
where $\tilde {\beta}=a^{D-4}/4g^{2}$. The representation R of U(N) can
be labelled by N
integers $r= (n_{1},n_{2},...,n_{N})$ with $n_{1}\geq n_{2}\geq ...\geq
n_{n}$ , where $n_{i}$	corresponds to the number of boxes in the i-th row of a
Young tableau.
The explicit expression  for the character $\chi _{R}(U)$ in terms
of the eigenvalues $e^{i\phi _{k}}$, $0\leq\phi _{k}\leq 2\pi$, of matrix
$U$ is given by the Weyl formula  \cite {Weyl}
\begin {equation} 
							  \label {m.5}
\chi _{(n_{1},n_{2},...,n_{N})}(U)=\frac{det [e^{i(N-j+n_{j})\phi _{k}}]}
{det [e^{i(N-j)\phi _{k}}]},
\end   {equation} 
where the quantities insides the $\det$ are to be viewed as the $jk$ elements
of a matrix $M_{ik}$. However, for our purpose we need the explicit
dependence on the
matrix $U$ rather than on its eigenvalues. Such formulas are given by the
well-known Frobenius relation \cite {Weyl} between characters and symmetric
polynomials,
\begin {equation} 
							  \label {F}
\chi _{R}(U)= \sum _{\sigma \in S_{n}}\frac{\chi _{R}(\sigma )}{n!}
\prod _{j=1}^{K_{\sigma }}\tr (U)^{k_{j}} ,
\end   {equation} 
where $n$ is a number of boxes in the Young tableau corresponding to the
representation R of SU(N), $S_{n}$ is the symmetric group on n objects.
For a given permutation $\sigma $ $K_{\sigma }$ is a total number of cycles
and $\{k_{1},...  k_{j},... k_{K_{\sigma }} \}$  are cycle lengths. In the
right-hand side of (\ref {F}) $\tr U$  is trace in the fundamental
representation.
For the first characters one find explicit formula in \cite {Bars,Sam}.

The integrals $Z_{R}$ (\ref {m.4})
for the characters corresponding to the single
row Young tableau with n boxes can be  written in the form
\begin {equation} 
							  \label {m.7b}
Z_{(n,0,0,...)}=\int dU\int _{0}^{2\pi} \frac{d\phi}{2\pi}\frac{e^{-in\phi}}
{\det (1-e^{i\phi}U)}e ^{N\tilde {\beta}\tr (U+U^{+}-2)}.
\end   {equation} 
$Z_{0}\equiv Z_{(0,...)}$ has the  representation as a determinant of Bessel
function   \cite {Bars}
\begin {equation} 
							  \label {m.8}
Z_{0}(N\tilde {\beta},N)=e^{-2N^{2}\tilde {\beta}}
\det |I_{i-j}(2N\tilde {\beta})| .
\end   {equation} 
$Z_{0}(N\tilde {\beta},N)$ is nothing but the partition function of the
Gross and Witten model	\cite {GW}  and it exhibits the phase transition
\begin {equation} 
							  \label {m.10}
Z_{0}(N\tilde {\beta},N) \ninfty{\sim} \left \{
\begin{array}{ll}
\exp \{-N^{2}(\frac{3}{4}+ \frac{1}{2}\ln 2 \tilde {\beta})\}&
\mbox {if $\tilde {\beta} >1/2$}\\
\exp \{-N^{2}\tilde {\beta}(2-\tilde {\beta})\}&
\mbox {if $\tilde {\beta} \leq 1/2$}
\end{array}
\right .
\end   {equation} 
$Z_{(1,0,0...)}(N\tilde {\beta},N)=\frac{Z_{0}}{N}<\tr U>$ ,
can be found from the relation
\begin {equation} 
							  \label {m.9}
<\tr U>=N+\frac{1}{2N}\frac{\partial}{\partial \tilde {\beta}}
\log Z_{0}(N\tilde {\beta},N),
\end   {equation} 
where
\begin {equation} 
							  \label {m.10a}
<f>=\frac{\int dU f(U)\exp \{N\tilde {\beta}\tr (U+U^{+}-2)\}}
{\int dU \exp \{N\tilde {\beta}\tr (U+U^{+}-2)\}},
\end   {equation} 
 We have
\begin {equation} 
							  \label {m.15}
\frac{1}{N}<\tr U>=f(\tilde {\beta}),
\end   {equation} 
where
\begin {equation} 
							  \label {m.16}
f(\tilde {\beta}) \ninfty{\sim} \left \{
\begin{array}{ll}
1-\frac{1}{4\tilde {\beta}}&
\mbox {if $\tilde {\beta} >1/2$}\\
\tilde {\beta}&\mbox {if $\tilde {\beta} \leq 1/2$}
\end{array}
\right .
\end   {equation} 

The  explicit representations for the first $Z_{R}$'s  can be found from the
explicit representations (\ref {m.7})-(\ref {m.6c}) and  the generating
functional
\begin {equation} 
							  \label {m.11}
Z(A, A^{+})=\int dU e ^{N\tilde {\beta}\tr (UA^{+}+U^{+}A)}
\end   {equation} 

The integral (\ref {m.11}) defines the well known object,
this is  the partition function of the Brezin-Gross model which describes
one link gauge field in the external matrix source \cite {BrG}.

Let us examine the behaviour of $Z_{R}$ in the large N limit.
Substituting the
Frobenious relation (\ref {F}) for the characters (\ref {F}) in (\ref {m.4}),
\begin {equation} 
							  \label {m.17}
{\cal Z}_{R_{Y_{n}}}=
\sum _{\sigma \in S_{n}}\frac{\chi _{R}(\sigma )}{n!}\int
\prod _{j=1}^{K_{\sigma }}\tr (U)^{k_{j}}\exp \{N\tilde {\beta}
\tr (U+U^{+}-2)\}dU
\end   {equation} 
we see that different
permutations $\sigma$ give different powers of N in the integral over U.
The main
contribution comes from $\sigma$  with the most number of cycles, i.e.
$k_{i}=1$ and $K_{\sigma }=n$,
\begin {equation} 
							  \label {m.18}
{\cal Z}_{R_{Y_{n}}} =
\frac{\chi _{R_{Y_{n}}}((1^{n}))}{n!}
Z_{0}(f(\tilde {\beta})N)^{n}+{\cal O}(N^{n-1})
=\frac{d _{Y_{n}}}{n!}Z_{0}(f(\tilde {\beta})N)^{n}+{\cal O}(N^{n-1})
\end   {equation} 
where $d _{Y_{n}}$ is the dimension of the representation of symmetry group
given by the Young tableau $Y_{n}$ and	the factorisation property of
correlation functions in the large N limit
is taken into account. Note, that we used the Frobenius relation for
$SU(N)$ group and equation (\ref {m.16}) for $U(N)$, since we expect that
this difference can be neglected in the large N limit.
$dim _{R}$ being the character of the unit matrix
also has the similar asymptotic behaviour for the large N
\begin {equation} 
							  \label {2.11}
dim_{R_{Y_{n}}}=\frac{d_{Y_{n}}}{n!}N^{n}
+{\cal O}(N^{n-1}),
\end   {equation} 
The explicit formula for $1/N$ corrections to (\ref {2.11}) can be written
in terms of the lengths $n_{i}$ of the Young tableau  \cite {GT}.

Therefore for $n$ fixed we get
\begin {equation} 
							  \label {m.19}
 Z_{R_{Y_{n}}} =
Z_{0}(f)^{n}+{\cal O}(N^{-1}) .
\end   {equation} 
Note that (\ref {m.19}) for large $\tilde {\beta}$, fixed $n$ and large N
is in agreement with well-known answer	\cite {DrZu}
$Z_{R}=1-C_{2}(R)/2N\tilde {\beta}$, where $C_{2}(R)$ is a quadratic Casimir.

Representing the sum over all representations  in (\ref {m.3}) as
\begin {equation} 
							  \label {m.20}
{\cal K}(U_{0},U_{L}) = \sum _{n} \sum _{R\in Y_{n}}dim_{R}(Z_{R})^{(L-1)}
\chi _{R}(U_{0}U^{+}_{L}),
\end   {equation} 
where the second sum in (\ref {m.20}) is taken over all representations
whose Young tableaux are in the set of Young tableaux with n boxes, we have
\begin {equation} 
							  \label {m.21}
{\cal K}(U_{0},U_{L}) =(Z_{0})^{L'} \sum _{n} (f)^{nL'}\sum _{R\in Y_{n}}
dim_{R}  ~
\chi _{R}(U_{0}U^{+}_{L}).
\end   {equation} 
Using the Frobenius relation once again we represent ${\cal K}(U_{0},U_{L}) $
for large N as
\begin {equation} 
							  \label {m.22}
{\cal K}(U_{0},U_{L}) = (Z_{0})^{L'}\sum _{n}\frac{f^{nL'}}{n!}
 \sum _{R\in Y_{n}}dim_{R}
\sum _{\sigma \in S_{n}}\chi _{R}(\sigma )
\prod _{j=1}^{K_{\sigma }}\tr (U_{0}U^{+}_{L})^{k_{j}},
\end   {equation} 
 From this formula it is clear that the effective action describing an
interaction
of the field living on two infinite two-dimensional planes has all powers on
$\tilde{U}_{0,z;i}\tilde{U}_{L,z;i}^{+}$.
If we are interested in the first term of the effective action which
is linear on $\tilde{U}_{0,z;i}$ we can keep in (\ref {m.22}) the
permutation with cycles $(1^{n})$ and we have
\begin {equation}
\label {m.23} {\cal
K}(U_{0},U_{L}) = (Z_{0})^{L'}\sum _{n}\frac{f^{nL'}}{n!} \sum _{R\in
Y_{n}} dim_{R}~\chi _{R}((1^{n}) ) (\tr [U_{0}U^{+}_{L}])^{n}+....  
\end{equation} 
Dots
denote the terms with higher power of $U_{0}U^{+}_{L}$.  Taking into
account (\ref {2.11}) at large N we have 
\begin {equation}
\label {m.25a} {\cal
K}(U_{0},U_{L}) = (Z_{0})^{L'}\sum _{n}\frac{f^{nL'}}{n!n!} \sum
_{Y_{n}}d^{2}_{Y_{n}} (\tr [U_{0}U^{+}_{L}])^{n}+..., 
\end{equation}
Recall that the sum
in (\ref {m.25a}) is taken over all Young tableaux with n boxes.  Due to
the relation
\begin {equation}
\label {m.24} \sum
_{Y_{n}} d^{2}_{Y_{n}}=n!
\end{equation}
we have for the
transition function 
\begin {equation}
\label {m.12} {\cal
K}(U_{0},U_{L}) =(Z_{0})^{L'}\exp \{f ^{L'}N\tr (U_{0}U^{+}_{L}) +...\}.
\end   {equation} 

Therefore for the large N and small $g^{2}$
the transition function can be represented as
\begin {equation} 
							  \label {m.14}
{\cal K}(U_{0},U_{L})=(2\tilde {\beta})^{-\frac{L'}{2N^{2}}}
\exp \{ -\frac{3}{4N^{2}}L'+
(1-\frac{1}{4\tilde {\beta}})^{L'}N\tr (U_{0}U^{+}_{L})+...\}.
\end   {equation} 
In the strong coupling regime we have
\begin {equation} 
							  \label {m.16a}
{\cal K}(U_{0},U_{L})=
\exp\{ -N^{2}L'\tilde {\beta}(2-\tilde {\beta})+
\tilde {\beta}^{L'}N\tr (U_{0}U^{+}_{L})+...\}.
\end   {equation} 
Substituting the representations for the transition function in (\ref {m.2})
we get
\begin {equation} 
							  \label {m.15b}
Z_{2+1}=(\tilde {Z_{0}})^{L'L^{2}}\int \exp \{(1-\frac{L'}{4\tilde {\beta}})N
\sum _{i,z}\tr [V_{z}U_{0,z;i}V^{+}_{z+i}\Omega _{z+i}U_{L,z;i}
\Omega ^{+}_{z}]\}
\end   {equation} 
$$\times \prod _{z}dV_{z}d\Omega _{z} (\prod _{i} dU_{0,z;i} dU_{L,z;i})
$$
for large $\tilde {\beta}$
and
\begin {equation} 
							  \label {m.15a}
Z_{2+1}=(\tilde {Z_{0}})^{L'L^{2}}\int \exp \{(\tilde {\beta})^{L'}N
\sum _{i,z}\tr [V_{z}U_{0,z;i}V^{+}_{z+i}\Omega _{z+i}U_{L,z;i}
\Omega ^{+}_{z}]\}
\end   {equation} 
$$\times \prod _{z}dV_{z}d\Omega _{z} (\prod _{i} dU_{0,z;i} dU_{L,z;i})
$$
for small $\tilde {\beta}$.   The constant $\tilde {Z_{0}}$ is different
in the different regimes.

It seems reasonable to assume that in the week coupling regime
\begin {equation} 
							  \label {m.17a}
L'=\log s   .
\end   {equation} 

Performing in the lattice effective action the limit $a \to 0$ one
immediately recognize an action similar to the action (\ref {1.ea})
with special form of matrix $M$,
\begin {equation} 
							  \label {m.18c}
M^{AB} = \left(\begin{array}{cc} 0 & \delta \\
\delta & 0 \end{array}\right), ~~\delta =1-\frac{1}{4\tilde {\beta}}\log s ,
\end   {equation} 
i.e.
\begin {equation} 
							  \label {m.18a}
S[V,\Omega,A_i] = (1-\frac{L'}{4\tilde {\beta}})N\int d^2 z
\tr [V(z)D^{-}_{i}V^{+}(z)\Omega _(z)D^{+}_{i}\Omega ^{+}(z)],
\end   {equation} 
where
\begin {equation} 
							  \label {m.18b}
D^{\pm}_{i}=\partial _{i}+A^{\pm}_{i},~~U_{L,z;i}=e^{aA^{+}_{i}(z)},
{}~~U_{0,z;i}=e^{aA^{-}_{i}(z)}
\end   {equation} 

In this example the analog of quark-quark scattering amplitude  is given
by the expectation value of two longitudinal Wilson lines
\begin {equation} 
							  \label {m.22a}
{\cal A}=\int d^{2}z e^{iqz}<{\cal V}(0){\cal V}^{+}(z)>
\end   {equation} 
\begin {equation} 
							  \label {m.23a}
{\cal V}(z)=\tr [P\exp \int dy A(y,z)]
\end   {equation} 
The lattice version of (\ref {m.23a}) is simply
\begin {equation} 
							  \label {m.24a}
{\cal V}_{z} =V_{0,z}V^{+}_{L,z}
\end   {equation} 

Note that in the model example of the  one-dimensional transversal plane one
can
perform the integration over fields $dV_{z}d\Omega _{z}$ explicitly. Indeed,
taking into account the form  (\ref {m.3})  for the transition function
${\cal K}(U_{0},U_{L})$
\begin {equation} 
							  \label {m.19a}
Z_{1+1}=\int \prod _{z =1}^{M}
{\cal K} (V_{z}U_{0,z}V^{+}_{z+i},\Omega _{z} U_{L,z}\Omega ^{+}_{z+i} )
dV_{z}d\Omega _{z} dU_{0,z;i} dU_{L,z;i}
\end   {equation} 
$$=\int \prod _{z =1}^{M}
\sum _{R}dim_{R}(Z_{R})^{(L-1)}\chi _{R}(V_{z}U_{0,z}V^{+}_{z+i}
\Omega _{z+i} U^{+}_{L,z}\Omega _{z}^{+})
dV_{z}d\Omega _{z} dU_{0,z;i} dU_{L,z;i}, ~~i=1,$$
and the orthogonality condition for characters
we have
\begin {equation} 
							  \label {m.20a}
Z_{1+1}=\int \sum _{R}(Z_{R})^{(L-1)(M-1)}
\chi _{R}(\Omega ^{+}V_{1} P_{+}\prod _{z} U_{0,z;i}V^{+}_{M}\Omega
 _{M}P_{-}\prod _{z} U^{+}_{L,z;i}) dU_{0,z;i} dU_{L,z;i}.
\end   {equation} 
Here $P_{+}$ and $ P_{+}$ are ordering and antiordering products.
This answer was obvious from the beginning since in this case we deal with
the two-dimensional QCD  \cite {Mig,Rou}.

\section{Concluding Remarks and Discussion}
Let us make some comments about the 4-dimensional case.
In this case the y-lattice is two-dimensional
and we have to deal with the following lattice integrals
\begin {equation} 
							  \label {d.1}
Z_{2+2}=\int \prod _{(z,i)}\int \exp \{ \frac{aN}{4g^{2}}
\sum _{y,\alpha}\tr (\tilde{U}_{y,z;i}\tilde{U}^{+}_{y+\alpha,z;i }
+\tilde{U}_{y+\alpha,z;i}\tilde{U}^{+}_{y,z;i }
-2)\}\end   {equation} 
$$\times\prod _{\begin{array}{c}
\tilde{U}_{y_{1},0,z;i}=V_{y_{1},0,z}
U_{y_{1},0,z;i}V^{+}_{y_{1},0,z+i},~\tilde{U}_{y_{1},L_{2},z;i}=
V_{y_{1},L_{2},z}U_{y_{1},L_{2},z;i}V^{+}_{y_{1},L_{2},z+i}
\\
\tilde{U}_{0,y_{2},z;i}=V_{0,y_{2},z}U_{0,y_{2},z;i}
V^{+}_{0,y_{2},z+i},~
\tilde{U}_{L_{1},y_{2},z;i}=V_{L_{1},y_{2},z}U_{L_{1},y_{2},z;i}
V^{+}_{L_{1},y_{2},z+i},
\end {array}
}d\tilde{U}_{y,z;i}$$
$$\times dU_{y_{1},0,z;i}dU_{y_{1},L_{2},z;i}
dU_{L_{1},y_{2},z;i}dU_{0,y_{2},z;i}
dV_{y_{1},0,z}dV_{y_{1},L_{1},z}dV_{L_{1},y_{2},z}dV_{0,y_{2},z}
$$
In (\ref {d.1}) we have the product of the transition functions of the
two-dimensional lattice chiral model.
If we assume the periodic boundary conditions in one of two longitudinal
direction, say $y_{2}$-direction,one can expect that for $L_{1}\gg L_{2}$
the main contribution comes from the one dimensional excitations, i.e.
\begin {equation} 
							  \label {d.2}
Z_{2+2}\sim \prod _{y_{2}}(Z_{1+2})=(\tilde {Z_{0}})^{vol_{4}}
\end   {equation} 
$$
\times\int \exp \{(1-\frac{\log s}{4\tilde {\beta}})N
\sum _{i,z,y_{2},}\tr [V_{y_{2},z}U_{0,y_{2},z;i}V^{+}_{y_{2},z+i}
\Omega _{y_{2},z+i}U_{L,y_{2},z;i}\Omega ^{+}_{y_{2},z}]\}
$$
$$\times\prod _{z}(dV_{y_{2},z}d\Omega _{y_{2},z} (\prod _{i} dU_{0,y_{2},z;i}
dU_{L,y_{2},z;i}))
$$

In the continuous version the contribution of long one-dimensional tubes
apparently is still described by the action (\ref {m.18a}), in spite of
the summation over $y_{2}$ in  (\ref {d.2}), i.e.
\begin {equation} 
							  \label {d.3}
Z_{2+2}\sim (\tilde {Z_{0}})^{vol_{4}}
\int \exp \{ (1-\frac{\log s}{4\tilde {\beta}})N\int d^2 z
\tr [V(z)D^{-}_{i}V^{+}(z)\Omega (z)D^{+}_{i}\Omega ^{+}(z) \}
\end   {equation} 
$$\times \prod _{z}dV(z)d\Omega (z)(\prod _{i=1,2}dA^{+}_{i}(z)dA^{-}_{i}(z)),
$$
Equation (\ref {d.2}) describes a rough approximation, however it gives
a rather acceptable physical picture. This picture is in agreement
with the Verlinde and Verlinde answer
which is in accordance with perturbation calculations \cite {ChWu} and
has the shock-wave \cite {swt'H,swV,swNorma}
semiclassical interpretation \cite {VV}.

Let us stress that the question about dominate contributions in (\ref {d.2})
or in others words the question about the behaviour of non-linear $\sigma $
model with non-trivial boundary conditions is rather complicated and without
doubt it is worthy more attention.

The Lagrangians (\ref {1.13}) and (\ref {1.6}) show a relation between the
4-dimensional Yang-Mills theory and the two-dimensional chiral model.
A relation was expected long time ego
and was motivated by the fact that the both theories have
dimensionless coupling constant and both are asymptotically free. Note also
that
some similarity was expected between the usual local formulation of the
two-dimensional chiral model and the loop formulation, i.e.  dynamics of long
tubes excitations in high-dimensional
Yang-Mills theory\cite {Pol,Ar}. The Verlinde and Verlinde
truncated action \cite {VV} together with the assumption (\ref {d.2}) make
this expected relation more tangible.

In concluding, it has been argued that consideration of
quantum fluctuations of one-dimensional excitations confirms the
two-dimensional picture of high-energy scattering in QCD.

$$~$$
{\bf ACKNOWLEDGMENT}
$$~$$
The author is grateful to G.Arutyunov and K.Zarembo for useful
discussions.
$$~$$

\end{document}